\documentstyle[preprint,aps]{revtex}
\begin {document}
\draft
\preprint{UCI-TR 94-41}
\title{Transverse Lepton Polarization in Polarized W Decays}
\author{Shao Liu\footnote{Electronic address: lius@uciph0.ps.uci.edu}
and
Myron Bander\footnote{Electronic address:  mbander@funth.ps.uci.edu}}
\address{
Department of Physics, University of California, Irvine, California
92717}

%\date{October\ \ \ 1994}
\maketitle
\begin{abstract}
Calculations of transverse polarization of leptons in the decay
$W\rightarrow l\nu$ with polarized $W$'s are presented. Planned
accelerators will produce enough $W$'s for observation of the Standard
Model contributions to this polarization.  One loop corrections to the
polarization are given; these are too small to be seen at presently
available $W$ sources. The exchange of Majorons will contribute to these
polarizations; these may provide limits on the couplings of these
particles to leptons.
\end{abstract}

\pacs{PACS numbers: 12.60.Cn, 13.38.Be, 13.88.+e}
%\narrowtext
\section{INTRODUCTION}\label{introduction}
Planned multi-TeV, high luminosity ${\bar p}p$ colliders will be a source
of a large number of $W^{\pm}$ bosons. At the LHC with a luminosity of
$10^{34}$ s${}^{-1}$cm${}^{-2}$ a cross section of $130$ nb
\cite{Wprodrate}, yields a $1300$ Hz production rate; the decay
$W\rightarrow l\nu$, for any lepton,  occurs 10\% of the time. Due to
the $V-A$ structure of the $W$ production vertex, low
$p_{\perp}$ $W$'s come out polarized \cite{Wpolar}.
Theoretical calculations of the polarizations involving the $W$ spin,
\mbox{\boldmath $J$}, and the transverse spin of the lepton,
\mbox{\boldmath $s$}, at
the $5\times 10^{-5}$ or even lower levels, in comparison with
experimental measurements
will provide tests of: (a) the Standard Model, (b)
radiative corrections to the Standard Model, and (c) new physics beyond
the Standard Model. For the latter we will look at the effects of
charged doublet Majorons \cite{Zee,BerSan,HHG} on the polarizations. As we
shall show, without any radiative corrections, in the case
$W^{-}\rightarrow \tau^{-}{\bar \nu}_{\tau}$ decay, the Standard Model
predicts \mbox{\boldmath $J\cdot s$} polarizations around 1\%; radiative
corrections modify these at the $10^{-3}$ level with a 10\% change as the
mass of the Higgs varying between $M_W$ and $4M_W$. These effects are
larger when the lepton momentum is parallel to the quantization axis for
the $W$ spin; unfortunately, due to angular momentum  conservation, the
rate is very small, and vanishes in the limit $m_l\rightarrow 0$.

We shall be more interested in triple correlation, \mbox{\boldmath
$s\cdot J\times Q$}, where \mbox{\boldmath $Q$} is the momentum
difference of the outgoing lepton and neutrino.
The polarizations in this correlation are proportional to the
imaginary parts of the decay amplitudes.
Radiative corrections in the Standard Model yield
polarizations at the order of $10^{-5}$; however, one loop correction
due to the exchange of charged Majorons can result in significantly
larger polarizations.

In Section \ref{polkin} we obtain various polarization expressions
in terms of the matrix elements $F$ and $G$ in $W$ decays.
In Section \ref{radcorr} radiative corrections are evaluated to one
loop, both in the Standard Model and
the Doublet Majoron Model. Results are summarized in Section
\ref{summary}.

\section{POLARIZATION DYNAMICS}\label{polkin}
In the case which neutrinos are left handed and massless,
the most general $W^-l^-{\bar\nu}_l$ vertex has the form
\begin{equation}
{\cal M}_\mu={\bar u}(p_l)\left(F\gamma_\mu+\frac{Gm_l}{M_W^2}Q_\mu \right)
   (1-\gamma_5)v(p_\nu)\, ,\label{vertex}
\end{equation}
where $Q=p_l-p_\nu$, and $F$, $G$ are dimensionless quantities to be
evaluated . To the lowest order, we have $F=g/(2{\sqrt 2})$,
$g=e/\sin\theta_W$ and $G=0$.
As will be shown below, the polarizations of leptons
depend only on $G/F$, thus the radiative corrections to
$F$ which are small can be ignored.
In any case $F$ is determined by the the rate of $W\rightarrow
l^-+{\bar\nu}_l$. In addition the corrections to $G$ are kept to the
lowest order in $m_l/M_W$. Since this factor already appears in the
definition of $G$ in Eq.~(\ref{vertex}), $m_l$ is set to zero except
for a photon exchange diagram (to be discussed below) which has a
dependence on $\ln (m_l/M_W)$.

The differential decay rate for a $W$ at rest is
\begin{eqnarray}
\frac{d\Gamma}{d\mbox{\boldmath $\hat{Q}$}}&=&
\frac{1}{256\pi^2M_W}\left\{F^{*}F\left[
-\mbox{\boldmath
$Q\cdot\epsilon^*Q\cdot\epsilon$}+M_W^2\mbox{\boldmath
$\epsilon\cdot\epsilon^*$}
-iM_W\mbox{\boldmath $Q\cdot\epsilon\times\epsilon*$}
\right. \right. \nonumber\\
&+&m_l\left (iQ\mbox{\boldmath $s\cdot\epsilon\times\epsilon^*$}
+\mbox{\boldmath $s\cdot\epsilon
Q\cdot\epsilon^*$}+\mbox{\boldmath
$Q\cdot\epsilon s\cdot\epsilon^*$}\right ) \Big]\\ \label{diffrate}
&-&\frac{FG^{*}m_l}{M_W}\mbox{\boldmath $Q\cdot\epsilon^*$}\left
[i\mbox{\boldmath
$Q\cdot s\times\epsilon$}+M_W\mbox{\boldmath $s\cdot\epsilon$}\right ]+
\frac{F^{*}Gm_l}{M_W}\mbox{\boldmath $Q\cdot\epsilon$}\left [-i\mbox{\boldmath
$Q\cdot s\times\epsilon^*$}+M_W\mbox{\boldmath
$s\cdot\epsilon^*$}\right ] \Big\}\, .\nonumber
\end{eqnarray}
In the above equation \mbox{\boldmath $s$} is the {\em transverse}
polarization of the lepton, and \mbox{\boldmath $\epsilon$} is the $W$
polarization vector; we have $\mbox{\boldmath $Q\cdot s$}=0$.

Measurable quantities of interest are:\\
(i) the decay rate of $W$'s with spin $+1$ (transversely polarized)
along the \mbox{\boldmath
$J$} direction ($\mbox{\boldmath $J$}^2=1$):
\begin{eqnarray}
\frac{d\Gamma_{J=1}}{d\mbox{\boldmath $\hat{Q}$}}&=&\frac{F^{*}F}{512\pi^2M_W}
(M_W-\mbox{\boldmath $Q\cdot
J$})^2 \Bigg [1+\frac{2m_l}{M_W-\mbox{\boldmath $Q\cdot J$}}\left
(1-Re\frac{G}{F}\right )\mbox{\boldmath $s\cdot J$}\nonumber\\
&-&Im\frac{G}{F}\frac{2m_l}{M_W(M_W-\mbox{\boldmath $Q\cdot
J$})}\mbox{\boldmath $Q\cdot s\times J$}\Bigg ]
\, ,\label{transrate}
\end{eqnarray}
(ii) a similar quantity for a $W$ with spin $0$ (longitudinally
polarized) along the \mbox{\boldmath
$J$} direction:
\begin{eqnarray}
\frac{d\Gamma_{J=0}}{d\mbox{\boldmath $\hat{Q}$}}&=&\frac{F^{*}F}{256\pi^2M_W}
\left [M_W^2-(\mbox{\boldmath $Q\cdot
J$})^2\right ] \Bigg [1+\frac{2m_l}{M_W^2-(\mbox{\boldmath $Q\cdot
J$})^2}\left (1-Re\frac{G}{F}\right )\mbox{\boldmath $Q\cdot J$}
\mbox{\boldmath $s\cdot J$}\nonumber\\
&-&Im\frac{G}{F}\frac{2m_l}{M_W(M_W^2-\left (\mbox{\boldmath $Q\cdot
J$})^2\right )}\mbox{\boldmath $Q\cdot J Q\cdot s\times J$}\Bigg ]\, .
\label{longrate} \end{eqnarray}
The asymmetry in two polarizations is the largest when
\mbox{\boldmath $Q$} and \mbox{\boldmath $J$} parallel;
however, angular momentum conservation forces these rates to vanish in these
directions. We shall
present details of the polarizations averaged over angels in the case
of $W$ spin $+1$ along the \mbox{\boldmath $J$} direction.
For the \mbox{\boldmath $\hat{s}$} in the \mbox{\boldmath $\hat{Q}$}\/-
\mbox{\boldmath $\hat{J}$} plane case we find
\begin{equation}
{\cal P}_{\parallel}=\frac{3\pi m_l}{8M_W}(1-Re\frac{G}{F})\, ;
\label{inplane}
\end{equation}
and for the \mbox{\boldmath $\hat{s}$} normal to the
\mbox{\boldmath $\hat{Q}$}\/- \mbox{\boldmath $\hat{J}$} plane case
\begin{equation}
{\cal P}_{\perp}=\frac{3\pi m_l}{8M_W}Im\frac{G}{F}\, .
\label{outplane}
\end{equation}
Longitudinal polarized $W$'s give a zero angular averaged
transverse polarization of leptons; restricted angular averaged
yield results similar to the ones above, Eq.~(\ref{inplane})  and
Eq.~(\ref{outplane}).

\section{Radiative Corrections}\label{radcorr}

\subsection{Standard Model Radiative Corrections}\label{stmodcorr}
As may be seen from Eq.~(\ref{transrate}) and Eq.~(\ref{longrate})
radiative corrections to $F$ are irrelevant in all the polarization
expressions.  Thus only radiative corrections to $G$ should be considered.
Standard Model one loop corrections to $G$ in
Eq.~(\ref{vertex}) are presented in Fig.~\ref{diagrams}.
Diagram by diagram they are ultraviolet and infrared finite.
The lepton mass is set to zero in all loop calculations,
except in the case of the photon exchange interaction,
Fig.~\ref{diagrams}c,
which has a logarithmic dependence on the lepton mass.
An overall factor of lepton mass already appearing in the
coefficient of $G$ in Eq.~\ref{vertex} is retained.
In the case of $W\rightarrow \tau\nu$, results for the
real part of $G$ are plotted in Fig.~\ref{Gtau}. The lowest order
contribution with plane polarization, Eq.~(\ref{inplane}), is
${\cal P}_{\parallel}=2.6\times10^{-2}$, and should be readily
measurable. The Standard Model radiative corrections are $\sim
5\times 10^{-3}$ times of the above values, changing by 5\% as
$M_H$ varies between $M_W$ and $4M_W$.
Detection of these corrections is beyond the capabilities of the
presently available $W$ sources.

The imaginary part of $G$ is independent of the Higgs mass, and
it is given by
\begin{equation}
Im\, G=0.025\frac{g^3}{16\pi^2}\, .
\end{equation}
The polarization transverse to the \mbox{\boldmath $\hat{Q}$}\/-
\mbox{\boldmath $\hat{J}$} plane (Eq.~\ref{outplane}) is
\begin{equation}
{\cal P}_{\perp}=4.5\times 10^{-6}\, ;\label{outplaneSM}
\end{equation}
This is, again, too small for detectability.

The results for the muon may be obtained from those for the
$\tau$ by replacing $Re\, G_{\mu}(M_H)=Re\,
G_{\tau}(M_H)+9.15\times 10^{-3}$ and $Im\, G_{\mu}(M_H)=Im\, G_{\tau}(M_H)$.

\subsection{Charged Majoron Exchange}
The small value of $Im\, G$ resulted from calculations in the Standard
Model permits us to use it in
looking for additional contributions from models
beyond the Standard Model. In this paper, we look specifically at
models which contain charged Majorons. A specific example
having this property is the Doublet Majoron Model. It contains
two charged Majorons, each carrying lepton number equal to 2 and with
masses $m_{1,2}$.  As our analysis cannot separate the contributions of
these two particles we shall parameterize the results in terms of one
charged Majoron with an average mass, $m_h$ \cite{BerSan}; this mass
will be assumed to be greater than $M_W$. The
interaction vertex of interest for such a particle is
\begin{equation}
{\cal L}_{hl\nu}=f_{ab}{\bar\nu^c_a}(1-\gamma_5)l_b h^+
+ \mbox{\rm h.c.}\, .\label{majoronvertex}
\end{equation}
In the above, $a,b$ label lepton families and $f_{ab}=-f_{ba}$
are, as yet, unrestricted couplings; a summation over lepton species is
implied.  We shall be interested only in the contribution to $Im\, G$ due
to this  interaction. The Feynman diagram of the interaction is shown
in Fig.~\ref{chmajex}, and the calculated result
is shown in Fig.~\ref{chmajImG}.

Currently limits exist \cite{BerSan} on certain combinations of these
coupling constants.
\begin{eqnarray}
|f_{e\tau}f_{\mu\tau}|&\le &10^{-4}\left (\frac{m_h}{M_W}\right )^2
\, ,\nonumber\\
|f_{e\mu}|^2&\le &10^{-4}\left (\frac{m_h}{M_W}\right )^2\,
.\label{preslims}
\end{eqnarray}
These results do not rule out sizable values for either $f_{e\tau}$ or
$f_{\mu\tau}$. For subsequent discussion we shall assume that it is the
latter that dominates and show how polarizations in $W$ decay can
restrict this parameter.

\section{Discussion}\label{summary}
Although the Standard Model one loop corrections to $G$
are too small to be
seen with $W$ fluxes from planned sources, ${\cal P}_{\perp}$ in
Eq.~(\ref{outplane}) will be observable for a range of parameters
in extended Standard Models that incorporate charged Majoron mesons.
The transverse polarization of $\tau$ in the results shown in
Fig.~\ref{chmajImG} is
\begin{equation}
{\cal P}_{\perp}=\left (|f_{e\tau}|^2+|f_{\mu\tau}|^2\right )\times
       \left\{ \begin{array}{lll}
                2.7\times 10^{-4} &{}&  \mbox{for $m_h=M_W$}\\
                2.4\times 10^{-6} &{}&  \mbox{for $m_h=4M_W$}\, .
                \end{array}
             \right. \label{chmajperp}
\end{equation}
With the polarization sensitivities discussed in
Sec.~\ref{introduction}, namely $5\times 10^{-5}$, we find that for
charged Majoron mass close to the $W$'s mass, the out of plane
polarization can restrict $|f_{\mu\tau}|\ge 0.4$ for $m_h\sim M_W$ and
$|f_{\mu\tau}|\ge 5$ for $m_h=4M_W$. In these models neutrino masses
are \cite{BerSan} of the order of $m_{\nu}\sim m_0f_{\mu\tau}$ with
$m_0\sim 10^{-2}$ eV. A positive result for the perpendicular
polarizations would imply, within this model, $m_{\nu_{\tau}}\ge
4\times 10^{-3}$ eV.
\section*{Acknowledgments}
We wish to thank Dr. A.\ Lankford for various discussions. This work
was supported in part by the National Science Foundation under Grant
PHY-9208386.
%\nobreak
%\newpage

\begin{figure}
\caption{One loop corrections to the standard mode amplitude. }\label{diagrams}
\end{figure}
\begin{figure}
\caption{Standard Model contributions to the real part of the amplitude
$G$ for $W\rightarrow \tau + \nu_\tau$.}\label{Gtau}
\end{figure}
\begin{figure}
\caption{Charged Majoron Exchange.}\label{chmajex}
\end{figure}
\begin{figure}
\caption{Charged Majoron exchange contribution to
$Im\,G$ for $W\rightarrow \tau + \nu_\tau$.}\label{chmajImG}
\end{figure}
%\narrowtext
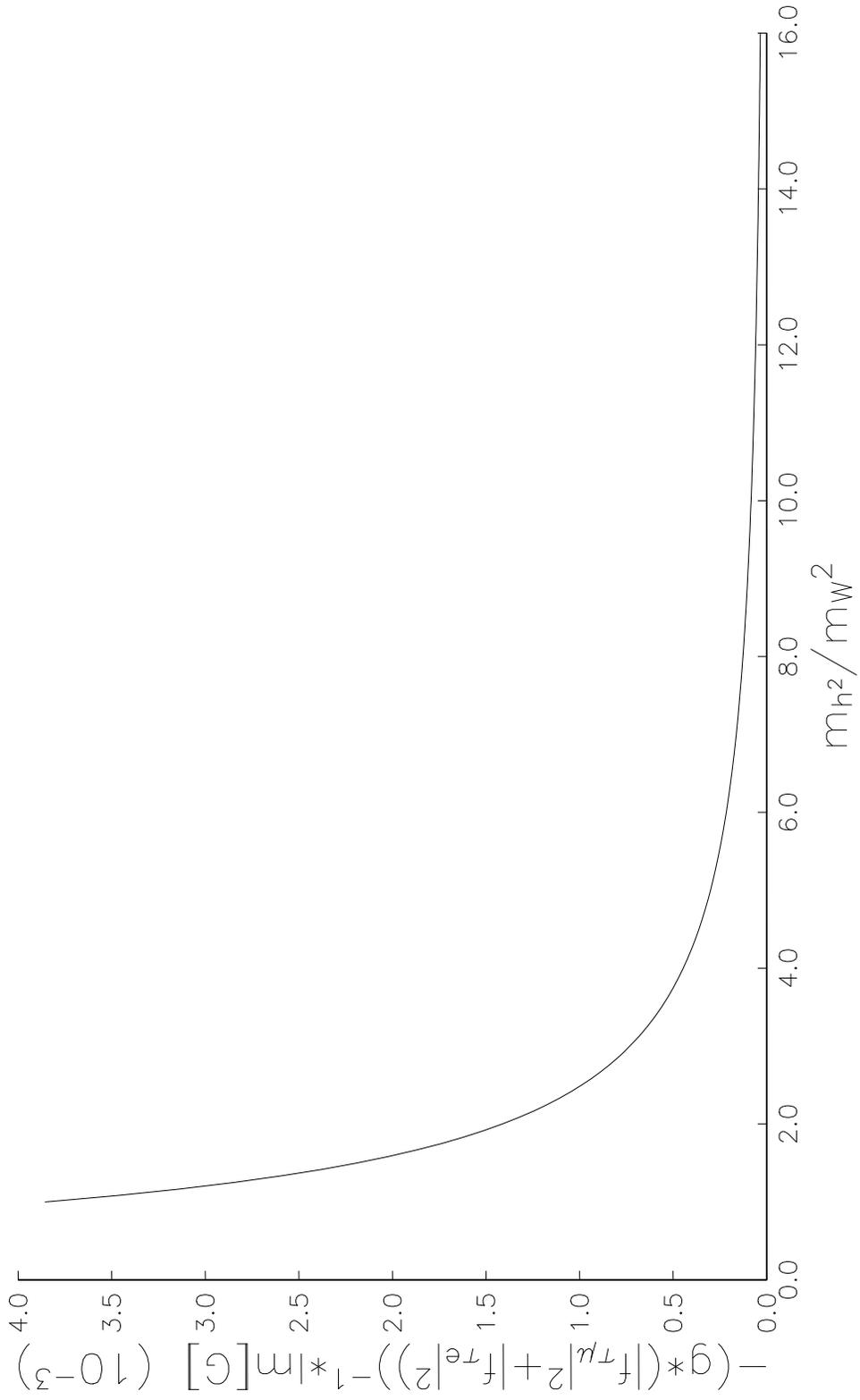
\end{document}